\documentclass[%
 reprint,
 amsmath,amssymb,
 aps,
]{revtex4-2}

\usepackage[version=4]{mhchem}
\usepackage{siunitx}
\usepackage{graphicx}
\usepackage{dcolumn}
\usepackage{bm}
\usepackage[hidelinks,colorlinks=true,linkcolor=blue,citecolor=blue]{hyperref}
\usepackage[english]{babel}
\begin{document}

\preprint{APS/123-QED}

\title{Energetics and dynamics of membrane necks in particle wrapping}

\author{Florent Fessler$^{1,2,3 \ \textcolor{blue}{*}}$ }
\author{Pierre Muller$^1$}
\author{Antonio Stocco$^{1}$}
\email{florentfessler@nyu.edu, stocco@unistra.fr}

\affiliation{$^1$Institut Charles Sadron, CNRS, UPR-22, 23 rue du Loess, 67200 Strasbourg, France}
\affiliation{$^2$Center for Soft Matter Research, Department of Physics, New York University, 726 Broadway Avenue, New York, NY 10003, United States}
\affiliation{$^3$Department of Chemistry, New York University, 29 Washington Place, New York, NY 10003, United States}

\begin{abstract}

Transport of microscopic objects across biological membranes usually involves membrane deformation to enclose the object followed by detachment of the engulfed particle. However, in artificial membranes, this last topological remodeling step is in many cases not spontaneous due to the elastic stability of the neck structure formed upon complete particle wrapping. In this work, we use optical trapping to induce the wrapping of a non-adhesive microsphere by the membrane of a giant lipid vesicle and investigate the energetics and dynamics of the resulting neck structure. We find that neck formation occurs as a result of membrane shape energy minimization under the application of external force. Remarkably, increasing membrane tension could reopen the neck and reverse the wrapping process. This process shows a clear hysteresis and a degree of reversibility. Neck cleavage and particle detachment into the vesicle’s interior could not be triggered in the range of our optical forces. Systematic studies on the thermal dynamics of wrapped particles allowed to establish that diffusion properties of the system are in agreement with a coupling of the particle motion with the neck structure, modeled as a solid inclusion within the membrane. Interestingly, the wrapped-particle dynamics exhibited a tension dependency, which can be described as the sum of several drag contributions.

\end{abstract}


\maketitle


\section{\label{sec:intro} INTRODUCTION}

Wrapping and budding events constitute a crucial step of many transport processes across biological membranes \cite{alberts2022mbc,bonifacino2004mechanisms,wrapping1}. These membrane remodeling processes involve the formation of a narrow neck structure connecting the bud to the mother membrane, which eventually undergoes fission to create a daughter vesicle \cite{SchmidSandraL1997CVFA,KozlovskyYonathan2003MFMf}. Extensive efforts in the last couple of decades have unraveled the physical principles of adhesion-driven wrapping of micro- and nanoparticles by model lipid membranes and vesicles. Theoretical \cite{Deserno2004,Bahrami2014,Bahrami2016,Lipowsky2012,Lipo1,Gompper,Lipowsky_1998} and experimental studies have established phase diagrams for adhesion-driven wrapping \cite{Spanke2020,Spanke2021,vutukuriwrap}. The neck shape and associated energetics was not considered in depth, the process being primarily dominated by adhesion energy.
A different regime of interaction arises when the particle and membrane have no particular affinity, yet the microparticle is capable of deforming the membrane thanks to an external or active self-propulsion force. A thorough understanding of the principles underlying such force-driven wrapping processes is still lacking, despite its relevance to systems involving biological microswimmers like bacteria \cite{intercell1,intercell2} and artificial ones like active Janus colloids \cite{Xiao2022,vutukuri2020active,Fessler2024}. Recent experiments have demonstrated that external forces, such as those provided by optical tweezers, can lead to stable fully wrapped configurations, even in the absence of significant adhesion energy \cite{Meinel2014,Fessler2023}. In this case, the stability of the fully wrapped state crucially depends on membrane shape energy, in particular at the neck, and membrane properties such as spontaneous curvature $m$ and tension $\sigma$ are expected to have a dramatic influence on the wrapping stability.\\

Additionally, in force-driven wrapping, due to the absence of strong attractive forces, the distances between the membrane and particle surfaces can be significantly larger than the few nanometer distance expected for the adhesive particle case. This difference impacts the particle dynamics on the membrane. A relatively large gap distance could lead to very different particle translational drag as compared to the ones measured for engulfed adhesive particles \cite{Pouligny1,Dimova1999,hamada,Kraftwrap}. Recent experiments for adhesive particles spontaneously  wrapped by vesicle membranes show that the particle translational drag depends not only on the membrane surface viscosity but also on the wrapping degree and 
 the finite distance between the particle and the membrane \cite{Kraftwrap,Marque}. Still, it is not clear how the drag is affected by the presence of the neck and to which extent it depends on the geometry of the system. While some experimental results were satisfactorily interpreted using models considering an inclusion either with a thickness comparable to the membrane \cite{Saffman1975} or protruding into the bulk fluid nearby \cite{DANOV199536,DanovKrassimirD.2000Vdoa,Fischer2006}, no predictions so far account for the exact geometry of a membrane neck connecting the mother membrane and wrapped particle. One of the reasons why such predictions are difficult to establish is because of the unknown detailed shape of the neck (due to the limited resolution when using optical methods), which should give the area of the membrane whose motion is coupled to the particle and predict the dominating dissipation modes. Moreover, additional confinement effects such as the presence of solid wall close to the membrane-particle system are also expected to impact the drag experienced by the particle.\\ 

Here, we investigate the stability and dynamics of membrane necks formed during microsphere engulfment by a lipid vesicle, driven by optical forces. Our model system features giant unilamellar vesicles (GUVs) with a small negative membrane spontaneous curvature induced by a sugar asymmetry across the lipid bilayer \cite{Gunther1999,Fessler2023,bhatia2} and tunable tension using a micropipette setup. We start by systematically varying the membrane tension, thereby probing the elastic stability of wrapping-induced neck structures. Next, we explore how wrapping affects the Brownian motion of the wrapped particle. Varying the particle size and the membrane tension, we identify the origins of enhanced dissipation in this system, which can be affected by the presence of a nearby solid wall. Finally, we evidence a dependence of the wrapped particle mobility on membrane tension, where an increase in membrane tension leads to a decreased mobility, and discuss the origin of this phenomenon.

\section{\label{sec:matmet} MATERIALS AND METHODS}

\textit{Chemicals and microparticles.} The phospholipids used are 1-palmitoyl-2-oleoyl-sn-glycero-3-phosphocholine (POPC) and fluorescent 1,2-dioleoyl-sn-glycero-3-phosphoethanolamine-N-(7-nitro-2-1,3-benzoxadiazol-4-yl) (NBD-PE) lipids purchased from Avanti Polar Lipids (Alabaster, AL, USA). Solid polyvinyl alcohol (PVA, molecular weight 145 000), chloroform, sucrose, D-glucose were obtained from Sigma-Aldrich (St. Louis, MO, USA). All chemicals were used as received. Commercial spherical SiO$_2$ particles were purchased from microParticles GmbH (Berlin, Germany) with radii $R_P=0.49,1.15$ and $\SI{2.13}{\micro \m}$ and nominal size coefficient of variation CV $<5\%$. Diluted particle solutions were prepared from the highly concentrated mother dispersion (5\% (w/v) aqueous suspension) by performing a tenfold dilution in a 0.5 mL Eppendorf tube. The particles were subsequently thoroughly cleaned by performing 3 cycles of successive centrifugation, removal of the supernatant, and redispersion in fresh MilliQ water. \\

\textit{Lipid vesicles formation.} Giant unilamellar vesicles (GUVs) were prepared using the PVA (polyvinyl alcohol) gel-assisted method as described in \cite{Weinberger2013}. A 5\% (w/v) PVA gel was prepared by dissolving PVA in MilliQ water, then uniformly spread (200 $\mu$L) onto cylindrical wells of radius $R_{\rm well}=\SI{4}{mm}$ and depth $d_{\rm well}=\SI{5}{mm}$ machined in a polytetrafluoroethylene (PTFE) plate and dried at 80°C for 45 minutes. Subsequently, 5 $\mu$L of a 99:1 molar mixture of POPC and NBD-PE lipids in chloroform (1 g/L) was deposited on the dried PVA gel, followed by vacuum desiccation for 15 minutes. The lipid-coated gel was hydrated with 200 $\mu$L of 150 mM sucrose solution and sealed for 2–3 hours to promote vesicle growth. The vesicles were collected and sedimented in a 150 mM glucose solution ($\SI{200}{\micro L}$ of vesicles in sucrose transferred in an Eppendorf tube containing 1 mL 150 mM glucose solution). The density mismatch between the internal sucrose and the external sucrose-glucose solution allowed gentle sedimentation without deforming the vesicles. \\

\textit{Micropipette fabrication and implementation.} The capillaries used to fabricate micropipettes are borosilicate glass capillaries of inner diameter 0.58 mm and 1 mm outer diameter (Ref. GC100-10 Harvard Apparatus). Pulling of micropipettes was performed using the Sutter Instruments Co. P-97 Pipette Puller. Microforging of the pulled capillaries was done using a homemade microforge setup consisting in a heated tungsten filament on which sodium tetraborate decahydrate was deposited. Microforging process resulted in the clean cutting of the cylindrical micropipette tip with the desired inner diameter suitable for GUVs manipulation $\SI{3}{\micro m}<R_{pip}<\SI{6}{\micro m}$. When the micropipette is carefully filled without introducing air bubbles, it can be connected to the hydrostatic setup using a PTFE tube. This setup allows hydrostatic pressure to be applied via the height difference between the water tank and the sample, which can be precisely controlled using a micrometric screw. \\

\textit{Optical trapping and microscopy.} The setup used for simultaneous imaging and optical trapping in this work is a modified OTKB Modular Optical tweezers (Thorlabs Inc., USA). It involves a 976 nm single mode laser diode whose temperature is precisely controlled by a TEC controller and thermistor to ensure a stable power output of the laser, guaranteeing a constant trapping force. A 100X  objective (Plan Fluorite Oil Immersion Objective, 1.3 Numerical aperture, 0.16 mm working distance, Nikon) is used both to trap the particles and image the sample using a camera. A mercury light source (Nikon C-HGFI Intensilight) with a blue filter is used as an excitation source for the fluorescent NBD probes in the sample. A translation stage (NanoMax 300 3-axis MAX311D, Thorlabs Inc.) connected to a set of piezo controllers (K-cubes KPC101, Thorlabs Inc.) was used in combination with an arbitrary waveform generator (Agilent 33521A) for controlled and precise displacement of the sample relative to the optical trap. 

\textit{Acquisition and particle tracking.}  We used a Hamamatsu Orca Flash-4 CMOS camera with a maximum detector size of 2048×2048 pixels ($\SI{0.065}{\micro \m}$/pix with the 100x objective) and a high temporal resolution of up to 1000 fps. The center of mass ($x_p$, $y_p$) of particles was tracked using the open source software Blender's motion tracking feature \cite{Blender}, which employs the Kanade-Lucas-Tomasi algorithm, optical flow techniques, and a Kalman Filter. Manual input and refinement tools ensured precise and reliable 2D tracking, even in challenging bright-field microscopy conditions or for particles with changing aspect due to position fluctuations around the focus plane. \\

\textit{Sample cell preparation.} The sample cell consists in two glass coverslips (Menzel Gl\"aser, Germany) separated by a self-adhesive silicone isolator purchased from Grace Bio-labs (OR, USA) containing a volume of solution $V\approx \SI{170}{\micro \L}$. The silicone isolator presents a $\approx 1$ mm wide slit to allow the micropipette to access the sample. In order to obtain deflated vesicles, one disperses a volume $V_{GUV}\approx1-\SI{2}{\micro \L} $ of concentrated vesicle solution in an open observation cell containing a volume $V_{sol}=\SI{180}{\micro \L}$ of matching concentration glucose solution and leave the sample open for the solvent to partially evaporate. Over the course of time, the evaporation of water will generate an osmotic imbalance, which leads to a slow deflation of the vesicles. A volume $V_{p}\approx1-\SI{2}{\micro \L} $ of diluted particles solution is also added to the sample. After about one hour, the sample is sealed (i.e. the top coverslip is put into place) to slow down the evaporation and allow correct optical visualization. Experiments are carried out in this configuration. Note that when using a micropipette, a small slit to allow insertion of the micropipette remains, making the sample not completely sealed.

\section{\label{sec:results} RESULTS AND DISCUSSION}

\subsection{\label{subsec:results1} NECK FORMATION}

\begin{figure}[t]
\includegraphics[width=0.7\linewidth]{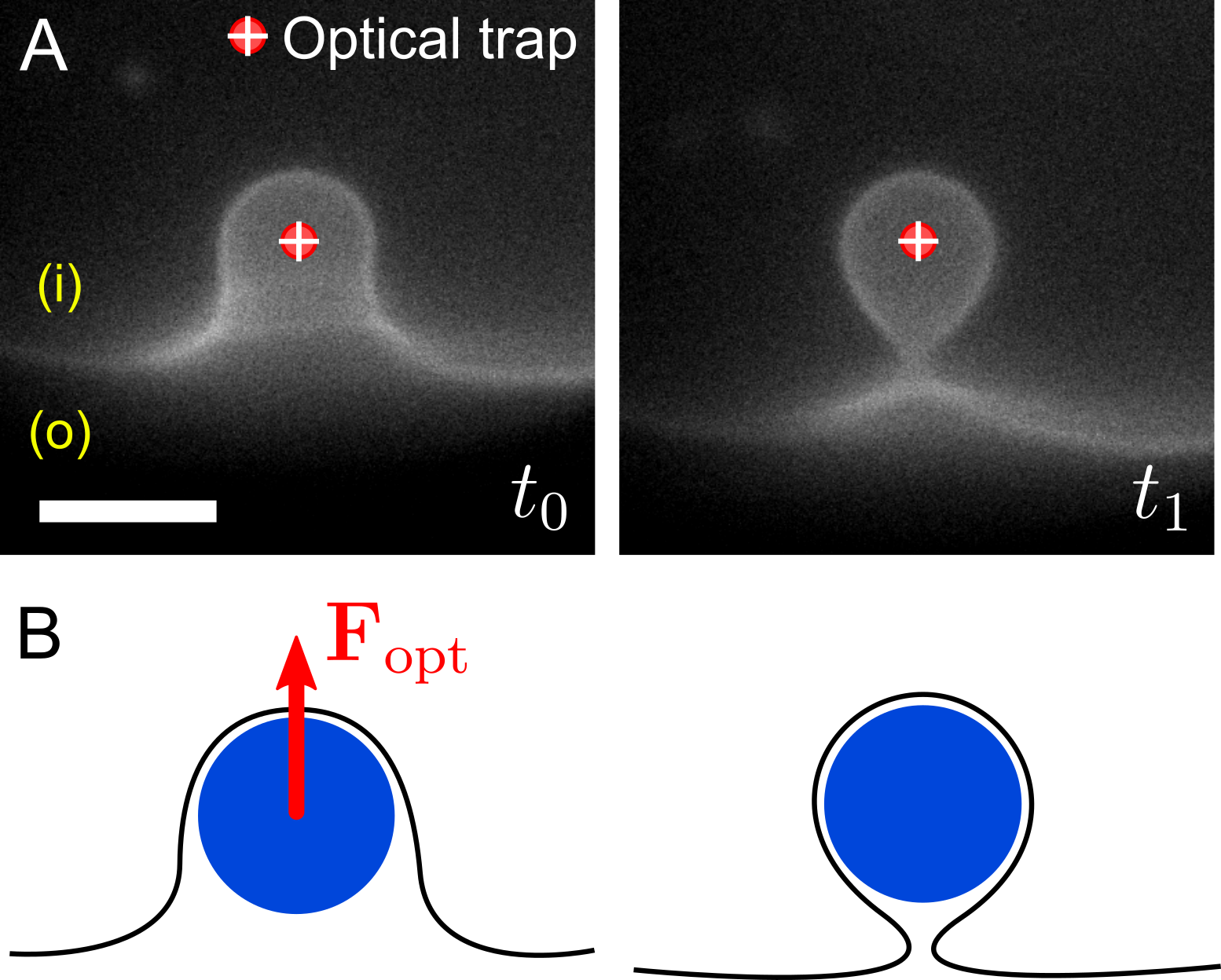}
\caption{
(A) Fluorescence microscopy snapshots of a neck formation event triggered by the relative pushing of a $R_P=\SI{2.13}{\micro \m}$ SiO$_2$ particle on a floppy POPC GUV membrane. (i) and (o) designate the GUV intravesicular and extravesicular space, respectively. The scale bar is $\SI{5}{\micro \m}$. (B) Schematic representation of the unbound membrane segment instability leading to the formation of a neck structure.}
\label{fig1}
\end{figure}

To experimentally investigate the energetics of wrapping-induced membrane necks, we study a model system consisting of a floppy ($\sigma<\SI{10}{\nano \newton \per \m}$) POPC giant unilamellar vesicle (enclosing a sucrose solution and suspended in a glucose solution) and a spherical SiO$_2$ microparticle. By imposing a relative motion between an optically trapped SiO$_2$ microsphere and a floppy GUV, we can bring the particle from a non-wrapping state to a stable full-wrapping state thanks to the force of the optical tweezers deforming the membrane \cite{Fessler2023}, as shown in Figure \ref{fig1} for a $R_P=\SI{1.15}{\micro \m}$ particle (see also Supp. movie S1\_MOV.avi). The fluorescence microscopy snapshots in Figure \ref{fig1}A together with the schematic representation in Figure \ref{fig1}B allow to visualize the neck closure process leading to the nucleation of an inward-pointing membrane bud containing the particle. This structure is always stable on the timescale of an experiment (up to several hours) even when the optical trap is switched off.

In contrast with adhesion-driven particle wrapping processes, the stability of the final full wrapping state in the forced particle wrapping experiment reported here cannot be understood considering only a competition between local membrane deformations (associated to the membrane segment bound to the particle) and adhesion \cite{Spanke2020,Spanke2021}. If very weakly attractive particle-membrane interactions can exist in the case of SiO$_2$ and POPC lipids, typical values for the adhesive energy per unit area $w$ are expected to be small with $| w| <\SI{0.1}{\micro\N \per \m}$ \cite{Gruhn2007,Fessler2024}. It is therefore readily evident that the energetic costs of local deformations induced by a spherical particle in the final fully wrapped state \cite{Deserno2004} outweigh the negative (favorable) adhesion energy contribution. This holds even in a low membrane tension regime and considering the existence of a weak negative membrane spontaneous curvature ($mR_P\ll1$), which is expected in our system due to the glucose/sucrose asymmetry across the membrane \cite{Fessler2023,Fessler2024}. Indeed, we have:

\begin{equation}\label{eq1}
    \mid w A_b\mid \ll  \sigma A_b + 8\pi \kappa_b (1+mR_P),
\end{equation}

where $\kappa_b$ is the bending rigidity, $m$ the membrane spontaneous curvature and $A_b=4\pi R_P^2$ corresponds to the excess membrane area that was pulled to the wrapping site and contributing to the adhesion free energy in the full engulfment state. This area exactly corresponds to the particle surface area considering the ideal case of a perfectly spherical final bud matching the particle size connected to a flat membrane by an infinitesimally small neck. In reality, the radius of the membrane bud enclosing the particle depends on the water gap separating the particle and the membrane, which is expected to be at least several tens of nanometers for nonadhesive particles. Still, Eq. \ref{eq1} predicts an unfavorable full wrapping state, due to the interplay between bound membrane segment deformations and adhesion. Hence,  the total free energy change $\Delta E>0$ between the non-wrapping and full wrapping states. It follows that the observed metastability should be provided by a local minimum of the full membrane shape energy. Bistability between non-wrapping and fully wrapped states was indeed predicted in the presence of a negative spontaneous curvature by full membrane shape energy calculations \cite{Jaime2021}. The bistability then results from the energetically costly shapes adopted by the unbound membrane segment for intermediate wrapping states, leading to an energy barrier for wrapping. This barrier is overcome by supplying external energy through optical tweezers, enabling the system to reach a metastable fully wrapped state.\\

Within the framework of spontaneous curvature elasticity theory, stability relations for membrane necks were derived by parameterizing the generic shape of vesicle-bud complexes as two hemispheres connected by two unduloid-shaped segments forming a neck of radius $r$. It can be shown that the bending energy of such a shape for a membrane with a spontaneous curvature $m$ has the form \cite{Lipowsky2022, Lipo1, Necks2Jaime}:

\begin{equation}\label{eqG}
    G_{be}=G_{be}(0) - 8 \pi \kappa_b\left( m-M_{ne} \right) r,
\end{equation}

where $G_{be}(0)$ is the energy associated to the shape far from the neck and $M_{ne}= \left(1/R_{ve}-1/R_s\right)/2$ the so-called closed neck curvature defined from the mean curvatures of adjacent segments of the mother vesicle and membrane bud with curvature radii $R_{ve}$ and $R_s$, respectively. It follows that for the neck structure to be stable, the free energy $G_{be}$ must increase with increasing $r$, leading to the stability criterion \cite{Lipowsky2022}:

\begin{equation}\label{stability}
    m \le M_{ne}.
\end{equation}

Calculating $M_{ne}$ for our case in the approximation $R_s\approx R_P=1.15 \ \mu$m and considering a size dispersion of the vesicles $7 \ \mu\rm{m} $ $< R_{ve}<20 \ \mu\rm{m}$ yields $ -3.6 \times 10^{5} \ \rm{m^{-1}}$ $> M_{ne}> -4.1 \times 10^{5} \ \rm{m^{-1}} $. Eq. \ref{stability} in turn implies $|m|>3.6 \times 10^{5} \ \rm{m^{-1}}$ as a lower bound for the spontaneous curvature absolute value in our system for it to be able to stabilize the formed neck. These values are consistent with spontaneous curvature values inferred in our experimental system \cite{Fessler2023} arising from a glucose/sucrose asymmetry across the lipid membrane \cite{Gunther1999}. The constrictions force at the neck $F_c=\frac{d G_{be}}{dr}=8\pi \kappa (m-M_{ne})$ are apparently not large enough to lead to spontaneous cleavage of the neck, contrarily to what was recently observed in the case of vesicles exposed to His-tagged proteins \cite{steinkuhler2020controlled}. Note that while this stability criterion seems to agree with our observations of neck stability for floppy vesicles, it does not take into account any influence of the membrane tension $\sigma$. Indeed, the latter could have a dramatic effect on the stability of neck, leading either to a re-opening or to a cleavage of the neck (fission). The influence of tension will be investigated in the next section. 

\subsection{\label{subsec:results2} NECK OPENING}

\begin{figure}[t]
\includegraphics[width=0.9\linewidth]{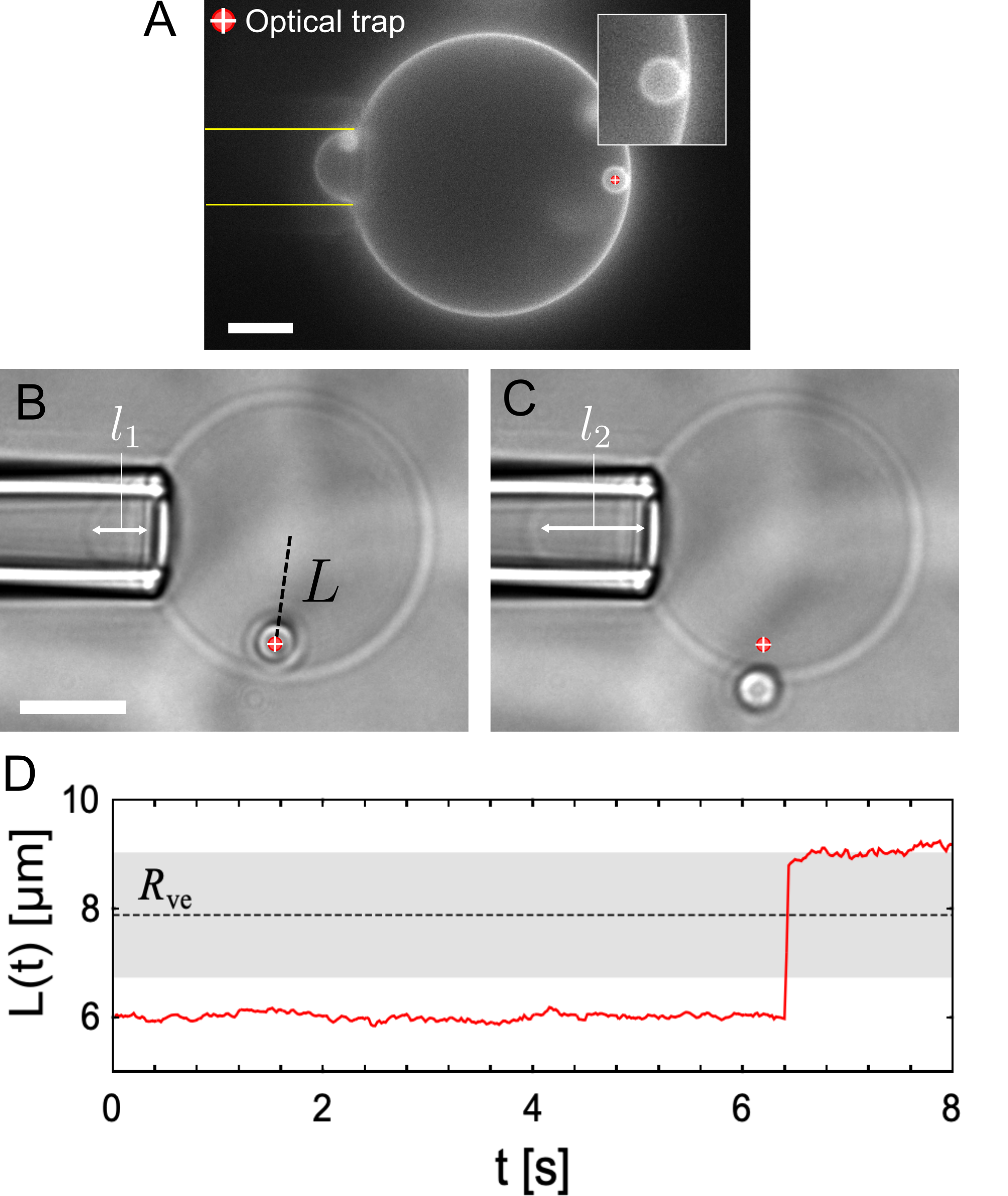}
\caption{Neck opening dynamics due to membrane tension increase for a $R_P=\SI{1.15}{\micro m}$ SiO$_2$ microparticle. (A) Fluorescence microscopy image of a POPC GUV aspirated with a micropipette (yellow lines) in which a SiO$_2$ microparticle was wrapped. The scale bar is $\SI{5}{\micro \m}$. The inset shows a 2× zoomed-in region containing the wrapped particle attached to the membrane. (B,C) Bright field microscopy images of a neck opening event, where the wrapped particle is being pushed by the membrane out of the optical trap, translating in an increase in the inter-center-of-mass distance $L$ and aspirated tongue length $l$ from $l_1$ to $l_2$. The scale bar is $\SI{5}{\micro \m}$. (D) Temporal evolution of the distance $L$ (as defined in (B)) where the sudden increase from $L< R_{ve}-R_P$ to $L\geq R_{ve}+R_P$ is a signature of the neck opening and subsequent particle expulsion.}
\label{fig2}
\end{figure}

Starting from the fully wrapped state, we aim at investigating the stability of the neck structure, thereby exploring whether particle endocytosis or particle unwrapping can be triggered. 
By optical tweezers, further forcing the particle towards the GUV center of mass leads to the nucleation of a membrane tube which never undergoes fission to complete particle endocytosis \cite{Fessler2023}. Inverting the relative motion direction between the wrapped particle and the GUV (distance between the particle and GUV center of mass is being increased), the particle escapes from the optical trap without membrane unwrapping in the accessible range of optical forces of our setup (see Supp. movie S1\_MOV.avi). This suggests that the neck formation introduces, if not an irreversibility, a hysteresis in the wrapping-unwrapping process. \\ 

Next, we use a micropipette to probe the influence of membrane tension on the neck by putting under tension the vesicle in which a particle was wrapped by the membrane. A fluorescent microscopy image of a typical experiment is shown in Figure\ref{fig2}A, where the wrapped particle is optically trapped with weak trapping stiffness at the vesicle equator to image both the particle and the aspirated membrane segment. The experimental protocol involves stepwise increments of the pressure difference in the micropipette $\Delta P$ using a hydrostatic device, which translates into membrane tension through Laplace's law. At each pressure step (i.e., membrane tension increase), we allow 1 minute for equilibration. In the meantime, the particle is maintained at the vesicle equator using a weak optical trap. There are two possible outcomes for such experiments. The first is depicted in Figure~\ref{fig2}B-C (and can be seen in Supp. movie S2\_MOV.avi), where the neck opens at a critical tension $\sigma_c$, resulting in a sudden radial displacement of the particle from the intravesicular space to the exterior. This displacement is evidenced by a stepwise increase in the particle-GUV center-of-mass distance $L(t)$, as shown in the plot in Figure \ref{fig2}D, transitioning from $L < R_{ve} - R_P$ to $L > R_{ve} + R_P$ in less than 100~ms. This corresponds to a lower bound for the particle escape velocity $v_{op} \geq \SI{20}{\micro \m \per \s}$. Simultaneously, an apparent area increase $\Delta A_{app}$ is observed, corresponding to the increase in length of the cylindrical portion $\Delta l = l_2 - l_1$ of the aspirated tongue (see Figure\ref{fig2}B-C for tongue lengths $l_1$ and $l_2$ definition). This area increase represents the surface area that previously wrapped the engulfed particle, confirming the opening of the membrane neck and the particle's release. We verify this by noting that $\Delta A_{app} = 2\pi R_{pip} \Delta l \approx 4\pi R_P^2$. Such an event is referred to as a \textit{neck opening} event, triggered by an increase in membrane tension. \\

\begin{figure}[b]
\includegraphics[width=1\linewidth]{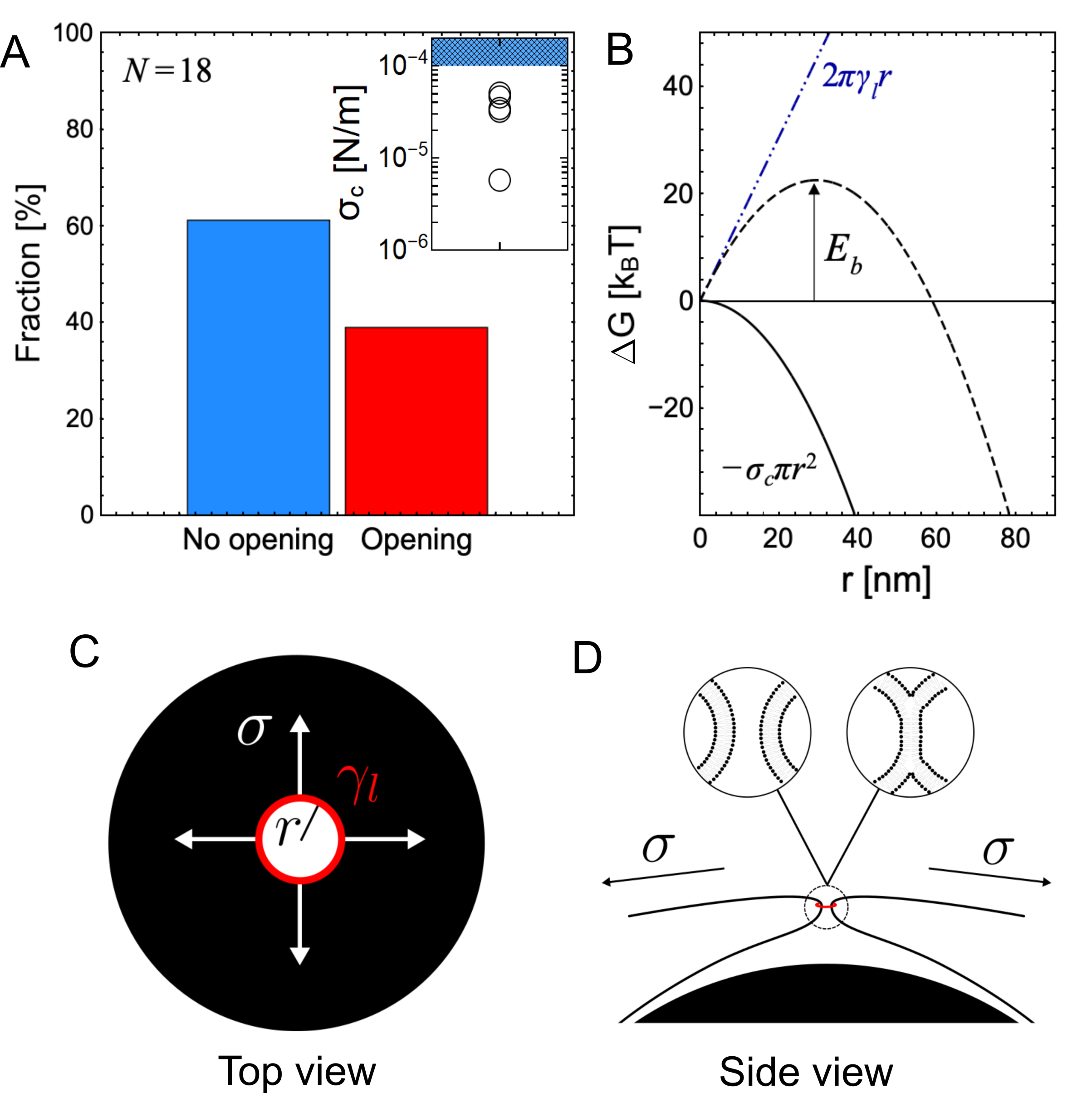}
\caption{(A) Neck opening statistics upon putting POPC vesicles with an engulfed $R_P=1.15 \ \mu$m Silica particle under tension for $N=18$ experiments. Inset shows the values of the critical tension $\sigma_c$ at which neck opening occurred when the critical tension could be accurately determined (for $N=5$ experiments on the total 7 opening events). The blue shaded zone corresponds to the membrane tension range that could not be probed with the experimental setup. (B) Free energy change $\Delta G$ versus neck radius $r$. The membrane tension contribution (plain black line), the line tension contribution (blue point-dash line) as well as the sum of the two contributions (dashed black line) are shown for $\sigma_c=34 \ \mu$N/m and $\gamma_l=1$ pN giving rise to an energy barrier with magnitude $E_b$. (C) Top view of the system illustrating the analogy with a pore where the membrane neck has a radius $r$ and a line tension $\gamma_l$ associated. (D) Side view of the neck profile with candidate membrane topologies at the neck.}
\label{fig3}
\end{figure}

 It occurs $\approx 39 $ \% of the observed cases. A critical tension $\sigma_c$ could be measured, and the mean value is reported in the inset of  Figure \ref{fig3}A and reads $\sigma_c = \SI{34}{\micro \newton \per \m}$.
The other outcome in our experiments is that the particle remains wrapped and no neck opening event occur in the range of pressure our setup allows to explore $\Delta P< 40  $ Pa, i.e. $\sigma<\SI{0.1}{\milli \newton \per \m} $. Data on the neck opening statistics for $N=18$ experiments are shown in Figure \ref{fig3}A, showing that in more than half of the cases ($\approx$ 61 \%), we could not observe an opening of the neck for the range of tensions accessible, pointing to a critical tension for neck opening $\sigma_c > \SI{0.1}{\milli \newton \per \m} $.\\

We note here that the stability criterion given by Eq.\ref{stability} does not account for any influence of the membrane tension $\sigma$. However, we just demonstrated experimentally that increasing the tension can in some cases lead to an unstable neck structure. In order to account for the contribution of tension, we make an analogy between our geometry of wrapped particle and treat the opening of the neck analogously to the opening of a pore, where the membrane tension $\sigma$ promotes the opening such that the change in total energy of the neck as a function of the neck radius $r$ is plotted in Fig \ref{fig3}B and reads \cite{LITSTER1975193}:

\begin{equation}\label{Gtot}
    \Delta G(r)=\Delta G_{be}(r)+ \Delta G_{\sigma}(r)=\gamma_l 2\pi r-\sigma \pi r^2,
\end{equation}

where $\Delta G_{be}\propto\gamma_l 2\pi r$ resists the wrapping and originates from Eq. \ref{eqG} where $\gamma_l=4\kappa_b (M_{ne}-m)$ is a line tension analogous to the one defined for pores, see schematics in Figure \ref{fig3}C-D. From Eq. \ref{Gtot}, the competition between the tension and bending energies lead to the existence of an energy barrier for neck opening $E_b=\pi \gamma_l^2/\sigma$. In our experiments when an opening was reported, the opening was spontaneous at the critical tension, meaning it was activated only by thermal energy such that $E_b(\sigma_c)=\pi \gamma_l^2/\sigma_c \lesssim 1-20\ k_B T$, see plot in Figure \ref{fig3}B. Following this reasoning, we obtain $\gamma_l\sim 1$ pN and $E_b$ not too large as compared to $k_B T$ using the experimentally measured critical tension $\sigma_c = \SI{34}{\micro \newton \per \m}$, as shown in Figure \ref{fig3}B. $\gamma_l\sim 1$ pN implies that $(M_{ne}-m)$ should be of the order of $10^{-6}$ m$^{-1}$ for the cases when neck opening could be triggered by the applied tensions. Using the experimental bound $M_{ne}=-3.6 \times 10^{5}$ m$^{-1}$, this yields $m\approx - 6\times 10^{-5}$ m$^{-1}$, which is in the expected range of $m$. These considerations point to the fact that the energetics of the neck in our system is well described in some cases by a competition of tension and curvature energy. However, for the cases when a tension of $\sigma \sim \SI{0.1}{\milli \newton \per \m}$ was not sufficient to trigger the opening of the neck (i.e. $\approx 61$ \% of the cases), it either means that the membranes in those cases possessed much larger spontaneous curvatures, or that other mechanisms than spontaneous curvature-induced constriction forces were at play. In particular, additional costs for opening are to be expected if a rearrangement of the lipids occurred at the neck, and a stalk or partial fusion of the membrane was initiated \cite{smirnova2021does}, as suggested by the schematics in Figure \ref{fig3}D. 

\subsection{\label{subsec:results3} WRAPPED PARTICLE DYNAMICS}

\begin{figure*}[ht!]
\includegraphics[width=0.88\linewidth]{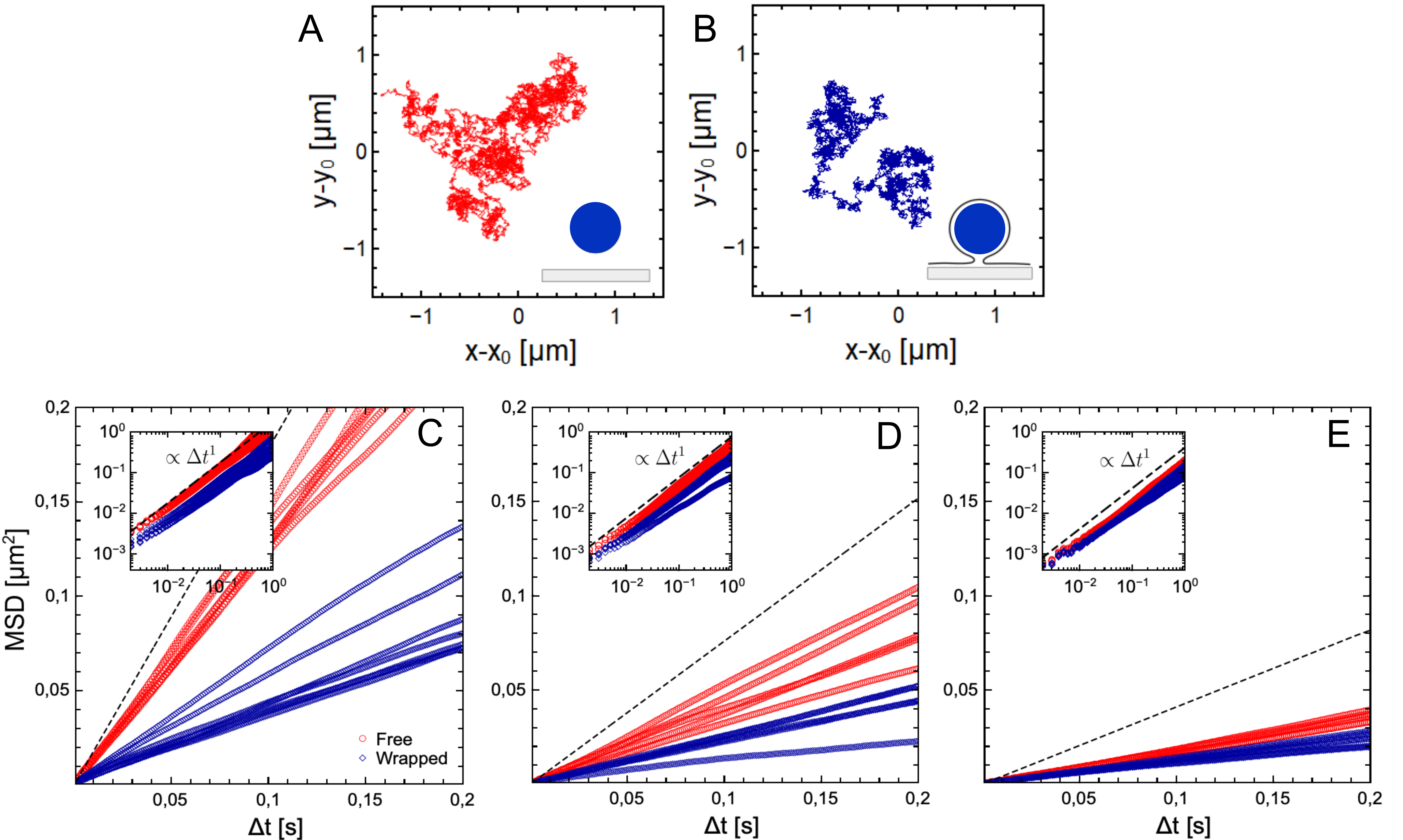}
\caption{Engulfed and free particle trajectories and two dimensional mean squared displacement curves (MSD). Representative trajectory in the $xy$ plane of a freely diffusing (A) and wrapped (B) $R_P=\SI{1.15}{\micro \m}$ microparticle. Schematics of the expected geometry is shown in inset (not to scale). (C-E) 2-dimensional mean squared displacement curves of free (red circles) and wrapped (blue diamonds) particles on short timescales for particles with radii $R_P=\SI{0.49}{\micro \m}$ (C), $R_P=\SI{1.15}{\micro \m}$ (D) and  $R_P=\SI{2.13}{\micro \m}$ (E). Inset shows a log-log representation up to $\Delta t=1$ s and evidences the power law ${\rm MSD} (\Delta t)\propto \Delta t^a$ with $a=1$. The dashed line stands for the predictions from Stokes-Einstein relation $D_{t,b}=k_B T/(6\pi \eta R_P)$. }
\label{fig4}
\end{figure*}

In the fully wrapped particle state, optical microscopy techniques do not allow to visualize the neck structure, given its small size (see e.g. fluorescence microscopy image in Figure \ref{fig2}A). In order to gain insight on the properties of the neck, we study the dynamics of the wrapped particle, which  should carry information on the neck structure and dynamics. To do so, we perform a set of particle tracking experiments on the Brownian motion of \ce{SiO2} particles with three different radii ($R_P=\SI{0.49}{\micro \m}$, $R_P=\SI{1.15}{\micro \m}$ and $R_P=\SI{2.13}{\micro \m}$) at an acquisition frequency close to 1 kHz (996 frames per second) in the free and fully wrapped situations. The so-called \textit{free} case refers to the situation where a particle diffuses close to the substrate (gravitational force confines the particle in a 2-dimensional plane close to the solid surface) in the absence of a lipid membrane in proximity, whereas the \textit{wrapped} case refers to the situation where the particle was previously fully wrapped in a lipid vesicle with optical tweezers, as described in Sec. \ref{subsec:results1}. Representative trajectories for free (in red) and wrapped (in blue) particles are shown in Figure \ref{fig4}A-B, with a total duration of the trajectory of $\approx 10$ s ($N=10 000$ points). \\

We plot the experimental two-dimensional mean squared displacement (MSD = $\left< (x_p(t+\Delta t)-x_p(t))^2 \right>+\left< (y_p(t+\Delta t)-y_p(t))^2 \right>$) curves associated to several wrapped and free trajectories ($N>5$ trajectories for each scenario) in Figure \ref{fig4}C-E. It appears that for all particle sizes, the MSD shows a purely diffusive behavior MSD $=4D_{t}\Delta t$ for $\Delta t< 1$ s, as shown in linear and log-log scale representation in inset. This confirms that the presence of a neck or tube does not introduce any elasticity at the considered timescales and agrees with the picture of a neck diffusing in the fluid lipid bilayer which only contributes as an additional dissipation of the thermal energy. Indeed, the diffusion appears systematically slowed down, i.e. linear MSD has a smaller slope, for the wrapped particle cases (blue diamonds) comparatively to the free cases (red circle), for all particle sizes.

\begin{figure*}[ht]
\includegraphics[width=1\linewidth]{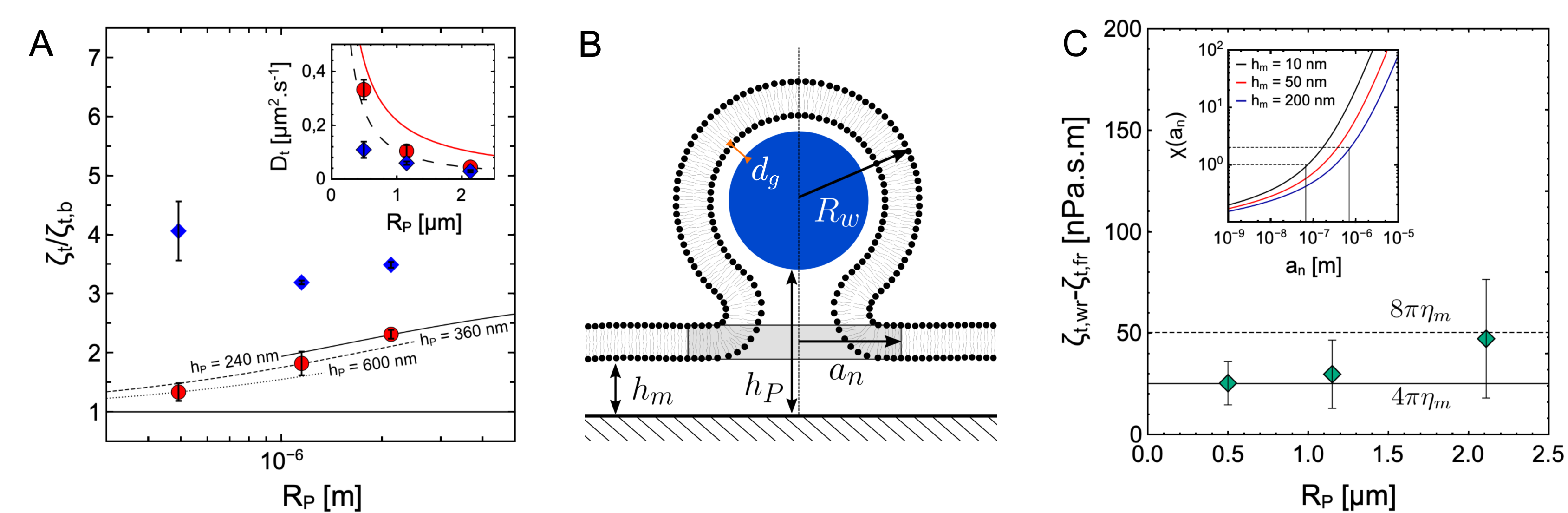}
\caption{Engulfed and free particle dynamics and associated drag. (A) Reduced experimental drag $\zeta_{t}/\zeta_{t,b}$ as a function of particle size for free (red discs) and wrapped (blue diamonds) cases with theoretical predictions from Faxén for three different gap distances. (B) Schematic representation of the wrapped particle situation with definition of the effective radius $R_w$, equivalent disk radius $a_n$, and film thickness $d_g$, membrane and particle gap distance $h_m$ and $h_P$, respectively. (C) Plot of the difference between the experimentally measured drags for wrapped and free particles $\zeta_{t,wr}(R_P)-\zeta_{t,fr}(R_P)$ as function of particle radius $R_P$. The plain and dashed lines are $4\pi \eta_m$ and $8\pi \eta_m$, respectively, which corresponds to $\zeta_{ES}$ as in Eq. with $\chi(\epsilon)=1$ and $\chi(\epsilon)=2$. Inset shows the function $\chi(a_n,h_m)$ as defined in Eq. \ref{Evans} where the dependence on $\epsilon$ is replaced by a dependency in the equivalent disk radius $a_n$ and membrane distance to the wall $h_m$, with $\eta=\eta_w=0.001$ Pa.s and $\eta_m=2 \times 10^{-9}$ Pa.s.m. 
}
\label{fig5}
\end{figure*}

To quantify this slowed-down diffusion, we extract the translational friction coefficient $\zeta_t=k_B T/D_t$ from individual trajectories by performing linear fits of the experimental MSD$(\Delta t)$ curves with ${\rm MSD}(\Delta t)=4D_{t}\Delta t$. We plot the average reduced friction $\zeta_{t}/\zeta_{t,b}$ as a function of particle size $R_P$ for both scenarios in Figure \ref{fig5}A, with $\zeta_{t,b}=6\pi \eta R_P$ the theoretical prediction for the friction coefficient in a bulk fluid from Stokes' law. In inset of Figure \ref{fig5}A, the diffusion coefficients $D_t(R_P)$ are also plotted. It appears that the experimental values of $\zeta_{t}(R_P)$ both in the free and wrapped cases are systematically greater than $\zeta_{t,b}(R_P)$ such that $\zeta_{t}/\zeta_{t,b}>1$ for all particle sizes. This increased friction is due to the presence of the underlying substrate imposing no-slip boundary conditions, leading to a particle-wall distance $h_P$-dependent (see Figure \ref{fig5}B) drag increase factor $f(h_P)$ such that $\zeta_t(h_P)=f(h_P)\zeta_{t,b}=f(h_P)6\pi \eta R_P$. Far from any interface, $f(h_P\gg R_P)=1$ for Stokes' law to be recovered. In the presence of a solid wall, an expression for $f(h_P)$ is given by the approximate solution to the Stokes equation derived by Faxén \cite{Faxen,BRENNER1961242}. The reduced drag $ \zeta_{t}(R_P, h_P)/\zeta_{t,b}$ for gap distances $h_P=240, 360$ and $600$ nm, were calculated \cite{Faxen,BRENNER1961242} and plotted in Figure \ref{fig5}B. These reduced drags can describe the friction experienced by free particles (red disks) with radii $R_P=\SI{0.49}{\micro \m} (h_P$ = 600 nm), $\SI{1.15}{\micro \m} (h_P$ = 360 nm) and $\SI{2.13}{\micro \m} (h_P$ = 240 nm), respectively. Such decreasing particle-wall distances with increasing particle size in the absence of lipid membrane agree with an equilibrium particle-wall distance $h_P$ determined by the balance of electrostatic and gravitational forces, as discussed in Supp. A. The monotonic increase of the reduced drag with increasing $R_P$ for free particles is therefore a result of the monotonic decrease of the particle-wall distance $h_P$ with $R_P$.\\

For the wrapped particle case (blue diamonds in Figure\ref{fig5}A), we measure a 3 to 4-fold translational drag increase from the bulk prediction value $3<\zeta_{t}/\zeta_{t,b}<4$. One can also note that in this case, $\zeta_{t}/\zeta_{t,b}$ is not monotonic with $R_P$. To rationalize the magnitude of this drag increase, we  start by accounting either (i) for a change in particle hydrodynamic radius $R_w>R_P$ of the wrapped particle if $h_P$ remains unvaried or (ii) to a change of the particle-wall distance $h_P$ if the radius remains the same (see Figure \ref{fig5}B for definition of $R_w$ and $h_P$). Doing so, we start by neglecting potential effects of the membrane viscosity. To describe our data, the hydrodynamic radius should strongly increase: $1.5<R_w/R_P<3$. This result is in contrast to fluorescence microscopy images of the wrapped particles, which provide a higher bound for $R_w$ such that $1<R_w/R_P<1.2$. Indeed, a small volume of fluid is enclosed in the membrane bud with the particle upon wrapping leading to the membrane bud radius being larger than the bare particle radius. However, it can be not larger than $R_P$ by more than a few hundreds of nanometers. Concerning the influence of the wrapping membrane on the equilibrium particle-wall distance $h_P$, it is hard to determine whether the wrapping has an effect for example on the electrostatic repulsion between the particle and the substrate, and in turn on $h_P$. Still, the presence of two bilayers between the particle surface and the solid wall in this geometry (see Figure \ref{fig5}B) allows to define a lower bound $h_P>10$ nm suggesting a higher bound value $f(h_P/R_P=0.01)\approx 3$ according to Fàxen and the expression for lubrication theory valid at such small gap distances \cite{BRENNER1961242}. Within this framework, considering the limiting scenario where $R_w=1.2 R_P$ and $h_P=10$ nm yields for the reduced translational drag:

\begin{equation}\label{modifstokes}
\frac{\zeta_{t,wr}}{\zeta_{t,b}}=\frac{f(h_P) 6\pi \eta R_w}{6\pi \eta R_P}=f(h_P)\frac{R_w}{R_P}< 3\times 1.2=3.6
\end{equation}

which could be one interpretation of the drag increase matching our measurements. However, this reasoning fails to capture the non-monotonic size-dependence of $\zeta_{t}/\zeta_{t,b}$ with $R_P$, suggesting a significant influence of the wrapping membrane and neck structure on the drag.  \\
 Now, we consider the fact that the connection between the wrapped particle and the membrane, in the form a neck, introduces an additional dissipation associated to the membrane 2-dimensional viscosity. The motion of the wrapped particle is indeed now coupled to the motion of this connecting structure within the lipid membrane, which experiences a drag expected to depend on the membrane viscosity $\eta_m$. While no model exists accounting for our complex geometry of a wrapped particle connected to a lipid membrane moving close to a wall, we can compare our results with some existing models describing ideal situations. In particular, we model the neck structure connecting the wrapped particle to the lipid membrane as a solid cylindrical inclusion with radius $a_n$ and account for the dissipations associated to its translational motion in the fluid lipid bilayer.\\

The translational drag on a cylindrical inclusion of radius $a$ and thickness comparable to the fluid membrane in which it is embedded was modelled by Saffman and Delbrück for a membrane with two dimensional viscosity $\eta_m$ \cite{Saffman1975}, and was later modified by Evans and Sackmann \cite{Evans1988} to account for the presence of a solid substrate closeby. This last model can be used in our case to estimate the additional contribution to the drag from a neck of radius $a_n$ that has to diffuse with the wrapped particle in the plane of the mother vesicle membrane (see scheme Figure \ref{fig5}B). In the case of a thin lubricating layer of thickness $h_m\ll \eta_m/\eta=\SI{1}{\micro \m}$ between the membrane and the underlying solid substrate, the drag force exerted on the translating disk is proportional to its velocity with the friction coefficient \cite{Evans1988}

\begin{equation}\label{Evans}
\zeta_{\rm ES}=4\pi \eta_m\chi(\epsilon), \ \  \chi(\epsilon) = \left[ \frac{1}{4}\epsilon^2  +  \frac{\epsilon K_1(\epsilon)}{K_0(\epsilon)}\right],
\end{equation}

where $\epsilon=a_n\sqrt{\frac{\eta}{h_m \eta_m}}$ is the dimensionless radius in the lubrication approximation and $K_1$ and $K_0$ are first and zero order modified Bessel functions of the second kind. Assuming that the wrapped particle-neck complex diffuses as a single solid object, we can sum the drag exerted on the disk inclusion with the modified Stokes drag from Eq. \ref{modifstokes}, which yields:

\begin{equation}\label{Evans2}
\frac{\zeta_{t,wr}}{\zeta_{t,b}}=\frac{f(h_P) 6\pi \eta R_w+\zeta_{\rm ES}}{6\pi \eta R_P}=\frac{R_w}{R_P} f(h_P)+\frac{2}{3}\frac{\eta_m}{\eta R_P} \chi(\epsilon)
\end{equation}

Here, the second term on the right hand side is proportional to the the Boussinesq number = $\eta_m/\left(\eta R_P\right)$ describing the relative importance of membrane viscosity contributions as compared to bulk viscosity, and introduces a $\propto 1/R_P$ dependency. The presence of this term could explain the larger measured drag for the smallest particle radius and the non-monotonic dependence with $R_P$. Indeed, we showed that the first term on the right hand side increases monotonically with $R_P$ due to the increase of $f(h,R_P)$ as the particle comes close to the substrate for larger particle size.\\ 

In order to confront these assumptions quantitatively, we plot in Figure \ref{fig5}C the difference $\zeta_{t,wr}-\zeta_{t,fr}$ of the experimentally measured drags as a function of $R_P$ (we subtract the measured drag value in the free case to the measured value for the wrapped case for each particle size). Doing so, we expect to isolate the contribution associated to the diffusion of the equivalent inclusion in the membrane $\zeta_{t,wr}-\zeta_{t,fr}=\zeta_{ES}$ (see Eq. \ref{Evans2}). It appears that the difference is almost constant with a small increase with increasing radius considering the error bars, and $4\pi \eta_m<\zeta_{t,wr}-\zeta_{t,fr}< 8\pi \eta_m$ for our range of particle size and using a typical value for POPC lipid membrane viscosity $\eta_m=2\times 10^{-9}$ Pa.s.m \cite{Faizi2022}. In other terms, our results agree with an almost constant contribution of $\zeta_{ES}$ given by Eq. \ref{Evans}, with $1<\chi({\epsilon})<2$. Note that in our system, while we do not know the radius of the equivalent disk that has to diffuse within the membrane $a_n$ nor do we know the membrane distance $h_m$, we can can assume $\eta$ and $\eta_m$ as they are tabulated properties of the bulk fluid and membrane, respectively. Hence, by plotting $\chi(a_n)$ for reasonable values of $h_m$ in inset of Figure \ref{fig5}C, we find that $1<\chi(a_n)<2$ implies $\SI{70}{\nano \m}<a_n<\SI{700}{\nano \m}$. Such values for the equivalent disk radius are realistic, despite being larger than the expected geometrical radius of the membrane neck considered so far (e.g. in Section \ref{subsec:results2}, where a closed neck would imply $r< 40$ nm). Other dissipative contributions could exist which could lead to the overestimation of the equivalent radius in this simple model of a particle coupled to a disk inclusion in a membrane, and these contributions will be investigated in the last section.

\subsection{\label{subsec:results4} TENSION-DEPENDENT DISSIPATION}

\begin{figure*}[t]
\includegraphics[width=0.95\linewidth]{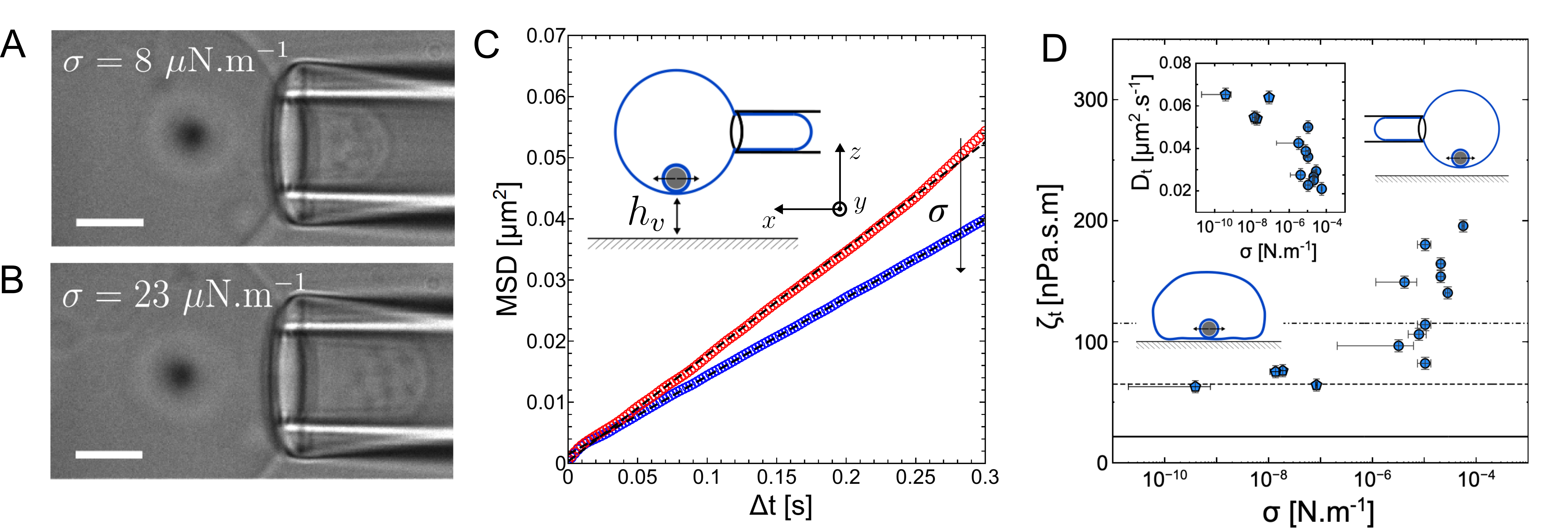}
\caption{ Bright field microscopy images of a GUV aspirated by a micropipette at (A) $\sigma=\SI{8}{\micro \newton \per \m}$ and (B) $\sigma=\SI{23}{\micro \newton \per \m}$, with an engulfed $R_P=\SI{1.15}{\micro \m}$ particle (out-of-focus) appearing as a blurry dark point. The scale bar is $\SI{5}{\micro \m}$. (C) One dimensional MSD along the y axis for an engulfed particle at two different membrane tensions. Red circles (top) corresponds to the lower tension membrane case while blue circles correspond to the tenser membrane case (bottom). Dashed black lines are  the fits yielding $D_{t}$ for the lower tension (red) and for the tenser (blue) vesicle. (D) Translational friction coefficient (with corresponding diffusion coefficients in inset) experienced by engulfed $R_P=\SI{1.15}{\micro \m}$ \ce{SiO2} particles as a function of vesicle membrane tension $\sigma$. Pentagons correspond to measurements where the vesicles were not under tension and the tension value was extracted from the force measurement. Disks correspond to measurements with vesicles under tension by the micropipette suction pressure (no optical trap). The sketches represent each situation. The solid black line accounts for Stokes' law in the bulk $\zeta_{t,b}$, the dashed line for the modified Stokes law with Faxén's factor $f(h_P)\approx 3$ at $h_P\approx 10 $ nm, i.e. $\zeta_t=3\times 6\pi \eta R_P$, and the dash-dotted line for $\zeta_t=3\times 6\pi \eta R_P+8\pi \eta_m$  (see Eq. \ref{Evans}, with $\chi(\epsilon)=2$).}
\label{fig6}
\end{figure*}

In the previous section, particle size was used as an experimentally tunable parameter to investigate the dissipative mechanisms in the specific geometry of an engulfed colloid connected to a floppy vesicle. Here, we explore the influence of a membrane property that already proved to have an influence on the neck in Section \ref{subsec:results2}: the membrane tension $\sigma$. \\

To do so, the experimental protocol again consists in performing long acquisitions at high frame rate of both the vesicle aspirated in the micropipette and the engulfed microparticle, as shown in Figure \ref{fig6}A-B. Due to the geometry and sedimentation of the particle, it is not possible to have both the particle and the aspirated vesicle segment in focus. We choose to keep the image focus on the aspirated membrane segment as it allows to make sure that the aspiration pressure is properly applied. The particle being out of focus can still be efficiently tracked with our tracking routine, and we do not expect it to introduce inaccuracies in our measurements of the particle position. Additionnally, geometrical corrections associated to the fact that the particle diffuses on a sphere (the vesicle, see sketch in Figure \ref{fig6}C) and not in a plane can be disregarded as the particle only explores a small region of space $A$ close to the bottom of the vesicle $R_{ve}$, such that $\sqrt{A}\ll R_{ve}$.\\

In Figure \ref{fig6}C, we report the one-dimensional MSD along $y$ of an engulfed particle upon increasing the membrane tension, corresponding to the images in Figure \ref{fig6}A-B. In a single acquisition of 30000 images acquired at 419 fps, the tension applied with the micropipette is increased step-wise from $\sigma=\SI{8}{\micro \newton \per \m}$ to $\sigma=\SI{23}{\micro \newton \per \m}$. The red curve shows the MSD for the particle motion during the first 10000 frames and the blue curve for the last 10000 frames (the motion is not analyzed during 10000-20000 frames as it corresponds to the time for the membrane to reach the higher equilibrium tension). It clearly appears that the slope of the MSD curve for lower tension is larger than for the higher tension case. Linear fits up to $\Delta t=0.3 $ s allow to extract $D_{t}=\SI{0.068}{\micro \m^2 \per \s}$ and $D_{t}=\SI{0.055}{\micro \m^2 \per \s}$ for the low and high membrane tension cases, respectively. MSD along $x$ is similar to the MSD along $y$ but shows oscillations arising from an external noise, see Supp. B. Such a relative increase in drag coefficient with increasing tension was systematically measured when reproducing this experiment in this geometry, also when using optical tweezers, see Supp. C. \\

 In order to extend the range of tensions probed, we measured the diffusion coefficient from MSD curves for particles engulfed in non-stressed floppy vesicles ($\sigma< 10^{-7}$ N.m$^{-1}$). The corresponding tension then is not the one applied with the micropipette, but the one measured from the tube force which contains information on the membrane tension. In Figure \ref{fig6}D, we plot the measured friction coefficient (and corresponding diffusion coefficient in inset) for these non-stressed low tension vesicles (pentagons) together with values measured from experiments where the membrane tension was controlled with the micropipette. Note that in the low tension regime (pentagons), the vesicle-substrate distance $h_v$ (defined in Figure \ref{fig6}C) can not be controlled as the vesicle is not manipulated with the micropipette. A significant contribution of the particle-wall proximity is expected, as already discussed previously. However, despite this contribution which is not present for higher tension as the vesicles are further away from the substrate, we can still observe a decrease in mobility for higher membrane tensions.

Quantitatively, the highest measured drag experienced by engulfed particles in the most tense vesicle is $\zeta_t=\SI{0.2}{\micro \Pa \s \m}\approx 10 \times 6\pi \eta R_P$,  which is approximately three times the drag measured for wrapped particles in floppy vesicles. Since none of the models introduced so far to interpret our drag measurements reveal an explicit dependence on the membrane tension, we will now discuss the possible origin of such high values of the measured drag for large $\sigma$.

 Deviations from the tension independent Saffman-Delbrück model \cite{Saffman1975,HStone1} were recently reported \cite{Domanov, Gambin2006, Quemeneur2014,HughesB.D.1981Ttar,NajiAli2007CttS}. In particular, the influence of membrane tension on the mobility of curvature-inducing proteins was evidenced and interpreted as an effect of the influence of tension on local protein-induced membrane deformations \cite{Quemeneur2014}. However, in this work, the sum of the considered contributions was shown to lead to the opposite dependency with tension, namely a decrease of the translational drag for increasing tension.

In our system, increasing tension leads to a change in the membrane deformation regimes. The bendocapillary length of the membrane $\lambda_{\sigma}=\sqrt{\kappa_b/ \sigma}$ varies from 3 µm at $\sigma$ = 10$^{-8}$ N.m$^{-1}$ to 30 nm at $\sigma$ = 10$^{-4}$ N.m$^{-1}$. Hence, the deformation induced by a particle with $R_P=\SI{1.15}{\micro \m}$ corresponds to a transition form a regime dominated by the bending $R_P<\lambda_{\sigma}$ to a regime dominated by the tension $R_P>\lambda_{\sigma}$\cite{Quemeneur2014}. The radius of the disk effectively diffusing within the membrane and attached to the wrapped particle can be regarded as the immobile part of the membrane subjected to significant curvature changes due to either the wrapping on the particle or to the formation of the neck.
Using the same analysis as in the previous section, a value of the disk radius $a_{n_2}\sim\SI{1}{\micro \m}$ gives a value of $\chi(a_{n_2},h_m)\approx4$ yielding the value of $\zeta_t\approx\SI{200}{\nano \pascal \s \m}$ measured on tense vesicles (Figure \ref{fig6}D).\\
As before, such large disk radius comparable to the particle radius is not only related to the neck size but to the immobile part of the membrane which is connected to the particle in its dynamics.  
In our geometry, a tension increase translates into a decrease of the distance between the mother vesicle membrane and the lipid bilayer wrapping the membrane. Indeed, for the low membrane tension regime, the interbilayer distance is expected to be large due to large membrane fluctuations and associated steric repulsion. A reasonable lower bound approximation for the interbilayer distance in this case would be the amplitude of the bilayer undulations at low tension $\xi_{\perp}\ge 200$ nm \cite{Bernard2000}. Higher membrane tension leads to suppression of these small wavenumber thermal undulations. In the absence of other repulsive forces such as electrostatics (lipids are zwitterionic), the interbilayer distance for high membrane tension can reach values as small as a few nanometers, until the hydration force prevents further thinning of the water gap. When interbilayer distance becomes small, shearing of the mother membrane upon translational motion of the wrapped particle may occur far from the neck location and starts only at relatively large interbilayer distances. However, the magnitude of these contributions in the case of a microparticle close to a tense membrane were reported to be 8\% greater at most from the bulk value at smallest accessible distances from the membrane using photonic force microscopy \cite{Rohr2015}.\\

It is worth noting that the high drag values measured for large membrane tensions correspond to cases where the neck did not open (\textit{no opening} cases), suggesting a higher spontaneous curvature of the membrane, as indicated by the results in section \ref{subsec:results2}. While no existing model explicitly predicts a dependence of membrane spontaneous curvature on the drag experienced by inclusions, its influence cannot be ruled out. Spontaneous curvature may directly affect either the size of the equivalent disk associated with the translating particle, which interacts with the membrane, or the membrane's effective viscosity.\\

Finally, an alternative interpretation of the high particle drag experienced at high tensions can be made considering that the leaflets constituting the membrane start to slide past each other due to the strongly curved membrane segments at the neck. This results into a intermonolayer friction contribution that becomes significant \cite{Quemeneur2014}. The intermonolayer friction coefficient in POPC membranes was measured to be of the order of $b\sim 10^9$ Pa s m$^{-1}$ \cite{intermono1,intermono2}. Typical associated friction contributions in our case of a membrane neck are expected to be of the order of $A_{i}b\approx \SI{100}{\nano \pascal \per \m}$ where $A_{i}= 100 $ nm$^{2}$ is a typical value of the contributing membrane area. Although it is not evident on which lengthscales such dissipation modes would extend, it appears that, taking the typical area associated to the neck $A_i\sim r^2= 100 $ nm$^{2}$ leads to a friction that could explain our friction data at high membrane tension Figure \ref{fig6}D.

\section{\label{sec:conclusion} CONCLUSION}

We demonstrated that nonadhesive microparticles can achieve stable wrapping by low-tension vesicle membranes under an external force, provided the membranes exhibit a small negative spontaneous curvature. The stability of the neck structure, crucial for maintaining the fully wrapped state in floppy GUVs, was explained using spontaneous curvature elasticity models. Through micropipette experiments, we assessed the reversibility of the force-driven wrapping process and quantified the membrane tension required to reopen the neck. The observed variability in this tension likely stems from differences in spontaneous membrane curvature or lipid reorganization at the neck.
Furthermore, we revealed that wrapped particles experience additional dissipation due to the coupling between their motion and the neck structure within the bilayer. Comparing experimental data for different particle sizes with theoretical models highlighted dissipative mechanisms tied to neck translation within the fluid membrane. Notably, we identified a strong dependence of particle mobility on the membrane's tension. This contrasts with prior reports for curvature-inducing proteins, which showed decreasing drag with increasing tension. Some potential dissipation pathways were identified, including effects involving intermonolayer friction. However, the results highlight the need for further theoretical and numerical efforts to fully account for this specific geometry.
These findings offer key insights into the physical principles governing the wrapping of particles by lipid vesicles or cells under directional forces or particle self-propulsion.


\begin{acknowledgments}
We acknowledge fundings from the Ecole Doctorale de Physique et Chimie-Physique from Université de Strasbourg and ANR EDEM (ANR-21-CE06-0042), and Andre Schroder for the help in implementing the micropipette setup.
\end{acknowledgments}


\section*{Supporting material}
\vspace{-0.1cm}
Movies S1\_MOV.avi, S2\_MOV.avi as well as a document containing 3 figures constitute the supplementary material of this work.


\section*{Author contributions}
F.F.: conceptualization, data curation, formal analysis, investigation, methodology, writing – original draft, writing – review \& editing. P.M.: investigation, methodology, writing – review \& editing. A. S.: conceptualization, funding acquisition, investigation, methodology, project administration, resources, validation, writing – review \& editing.

\vfill

\bibliography{apssamp}

\end{document}


\preprint{APS/123-QED}

\title{Supporting material: Energetics and dynamics of membrane necks in particle wrapping}

\author{Florent Fessler$^{1,2,3}$}
\author{Pierre Muller$^1$}
\author{Antonio Stocco$^1$}

\affiliation{$^1$Institut Charles Sadron, CNRS, UPR-22, 23 rue du Loess, 67200 Strasbourg, France}
\affiliation{$^2$Center for Soft Matter Research, Department of Physics, New York University, 726 Broadway Avenue, New York, NY 10003, United States}
\affiliation{$^3$Department of Chemistry, New York University, 29 Washington Place, New York, NY 10003, United States}


\keywords{Lipid membranes, giant vesicle, particle wrapping, membrane neck, optical tweezers, Brownian motion}

\maketitle


\section{Supplementary}

\subsection{Electrostatic and gravitational forces balance}\label{electrograv}

\begin{figure}[!b] 
    \centering
    \includegraphics[width=1\linewidth]{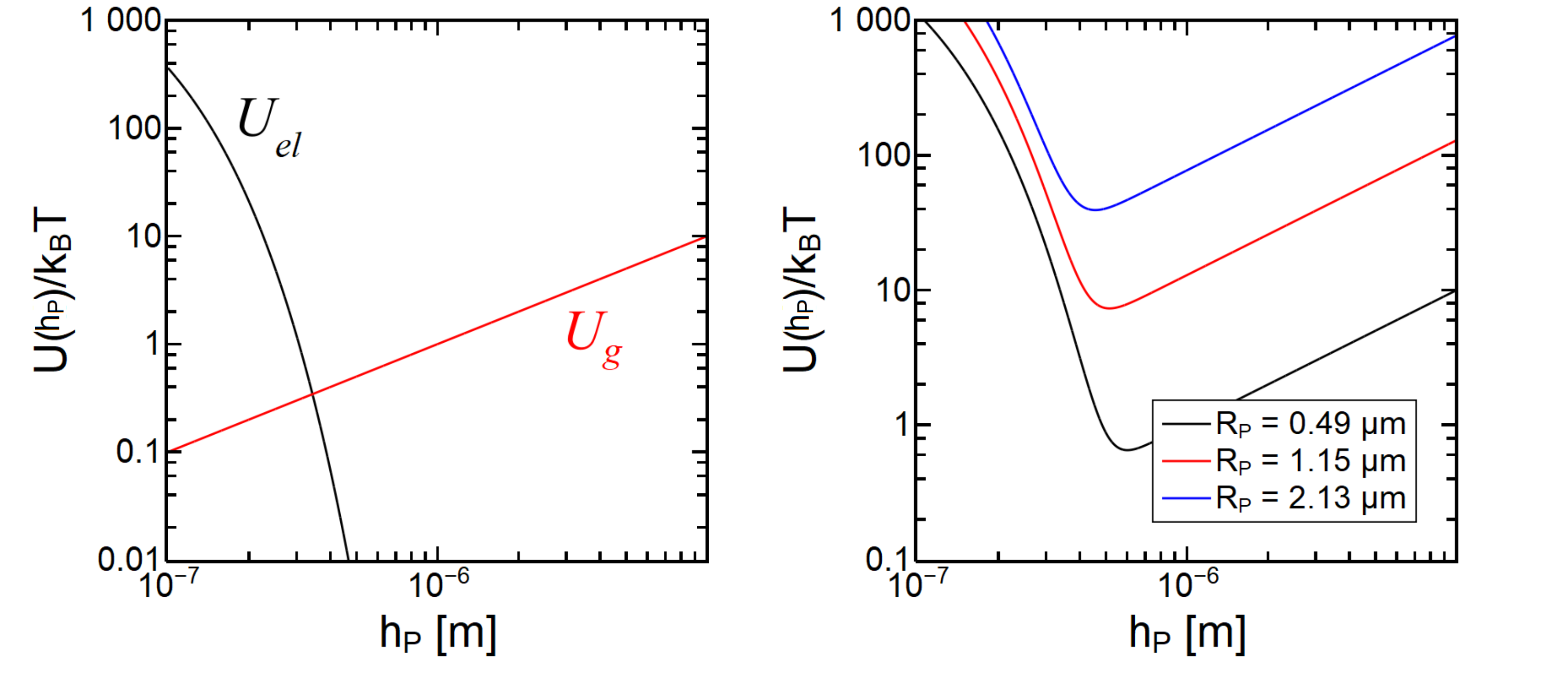}
    \caption{\textit{(left)} Distance dependent components of the potential $U_{el}$ for double layer repulsion and $U_g$ for gravity for a spherical SiO$_2$ particle with $R_P=\SI{0.49}{ \micro \m}$. \textit{(right)} Sum of the two components forming the total potential as in Eq. \ref{pottoteg} as a function of separation distance $h_P$ and for the three particle radii considered in Section IIIC (main text). We use here $Z_{part}=-\SI{25}{\m\V}$, $Z_{wall}=-\SI{75}{\m\V}$ and $\lambda_D=\SI{50}{\nano \m}$. Using representative values in this range ($Z_{wall}$ and $\lambda_D$ were not determined experimentally), one systematically obtains a minimum at smaller distance $h_P$ for larger particles.}
    \label{appendixelectro}
\end{figure}

At thermal equilibrium, a spherical SiO$_2$ bead immersed in a fluid will sediment at the bottom of the observation cell as it is denser than the surrounding fluid due to gravity. A particle may find a finite equilibrium distance from the substrate around which it might fluctuate due to thermal Brownian motion. This equilibrium distance results from the sum of the two main forces acting on the particle, namely gravity and the double layer electrostatic repulsion with the substrate. The latter repulsion is due to the particle and substrate surfaces being both negatively charged in water. The van der Waals interaction contributions are expected to be negligible in conditions where the gap $h_P$ between the particle and the substrate is relatively large \cite{flicker}. The total potential energy landscape therefore reads \cite{flicker,rashidi1,Ketzetzi2020}:

\begin{equation}\label{pottoteg}
\frac{U_{tot}(h_P)}{k_B T} = 
\begin{cases} 
B e^{-\frac{h_P}{\lambda_D}} + h_P \frac{\Delta \rho V g}{k_B T}, & \text{for } h_P > 0 \\ 
+\infty, & \text{for } h_P \leq 0 
\end{cases}
\end{equation}

\begin{figure}[!b]
\centering
\includegraphics[width=0.9\linewidth]{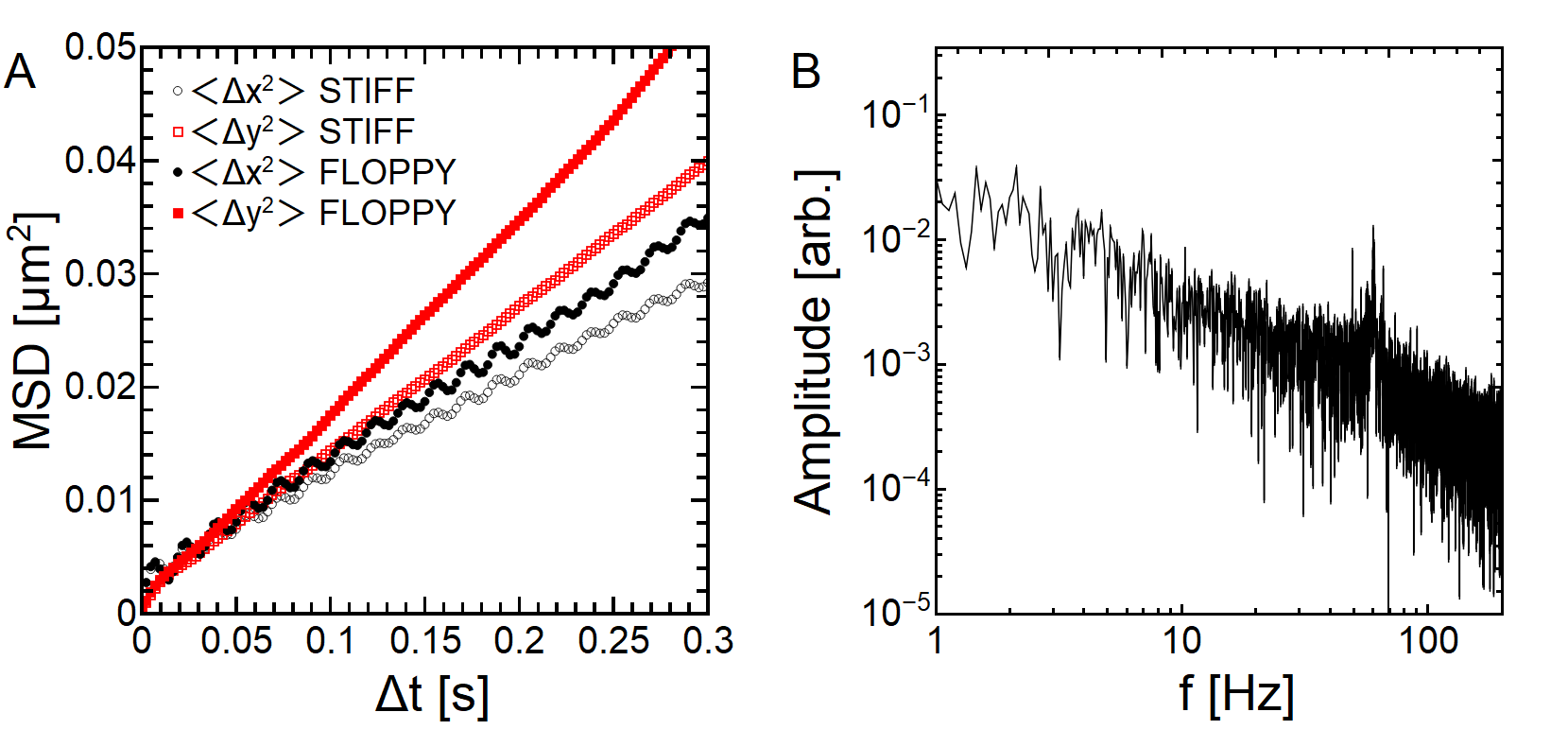} 
\caption{(A) Mean squared displacement curves along $x$ (black circles) and $y$ (red squares) showing both a decrease in slope upon tension increase, but with oscillations in the $x$ component of the movement. (B) Position fluctuation spectrum along $x$ associated with the corresponding trajectory for a less tense vesicle in (A), evidencing a peak around $f \approx 60$ Hz. This peak is reproducible when oscillations appear along $x$ and is therefore interpreted as vibrations of the optical table leading to fluctuations in the micropipette position along $x$.}
\label{oscill}
\end{figure}

where $\Delta \rho$ is the particle-fluid mass density difference and $V$ the volume of the spherical colloid $V=4/3\pi R_P^3$, $g$ the gravitational acceleration, the Debye length $\lambda_D$, and with:

\begin{equation}
    B = 64 \pi \epsilon_0 \epsilon_f R_P \left( \frac{k_B T}{e} \right)^2 \tanh \left( \frac{e Z_{\text{wall}}}{k_B T} \right) \tanh \left( \frac{e Z_{\text{part}}}{k_B T} \right),
\end{equation}

where $\epsilon_0$ is the vacuum permittivity, $\epsilon_f$ the relative dielectric constant of the fluid, $Z_{wall}$ the zeta potential of the substrate and $Z_{part}$ the zeta potential of the particle.\\

Note that the Debye length $\lambda_D$ can be written as:

\begin{equation}
    \lambda_D^{-1} = \sqrt{\frac{2 e^2 C_{\text{s}}}{\epsilon_0 \epsilon_f k_B T}}.
\end{equation}

where $C_s$ is the ionic strength. In absence of added salt (as in our case), this should only account for a contribution from the pH of the solution. In absence of salt, the Debye length should therefore be large $>\SI{100}{\nano \m}$, however due to impurities and CO$_2$ dissolution, it is not rare to have a smaller Debye length ($\lambda_D\approx \SI{50}{\nano \m}$) even in the absence of salt.\\

\begin{figure*}[!htb]
\centering
\includegraphics[width=0.9\linewidth]{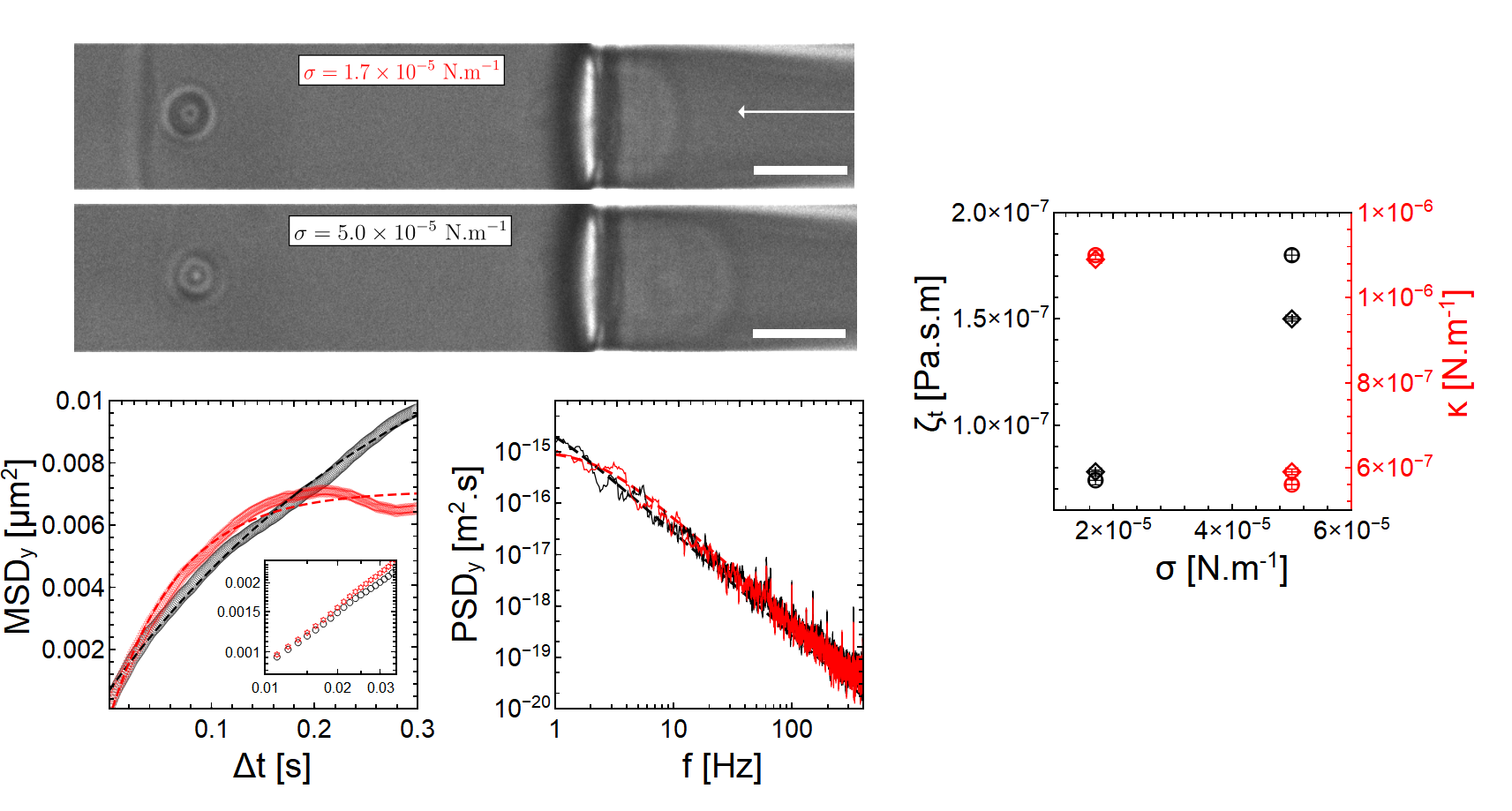}
\caption{\textit{(topleft)} Bright field microscopy images of the optically trapped engulfed particle at the equator of an aspirated vesicle for two applied tensions. \textit{(bottom left)} 1-dimensional MSD and PSD for the motion along the tangent axis to the vesicle membrane. Red and black dashed lines stand for fits with $\left< (y(t+\Delta t)-y(t))^2\right> = \frac{2k_BT}{\kappa}\left[1-e^{\frac{-\kappa}{\zeta_t} \Delta t}\right]$. for the MSD and $PSD(f)=\frac{D_t}{2\pi^2(f_t^2+f^2)}$ with $f_t=\kappa/2\pi\zeta_{t}$, for the PSD, with log-log scale representation of short time regime of the MSD in inset. \textit{(right)} Parameters extracted from the MSD (diamonds) and PSD (circles) fits as a function of applied membrane tension $\sigma$. Errorbars account for the fit standard deviation values. }
\label{tensiondeptrap}
\end{figure*}

Note that the electrostatic charge and Debye length in our system are not measured, but by using values used in the literature for similar systems \cite{Ketzetzi2020}, we can plot the potential as a function of the distance for each contribution $U_{el}$ and $U_g$ of $U_{tot}$ in Figure \ref{appendixelectro} \textit{(left)}. The total potential for different particle sizes is plotted in Figure \ref{appendixelectro} \textit{(right)}. We see that these energy considerations allow to explain the average distance $h_P$ from the wall of few hundreds of nanometers as inferred from Faxén's predictions using experimental results. Additionally, the fact that larger particles have smaller equilibrium distance $h_P$ is also captured by these considerations. However, more precise determination of the distance $h_P$ using these theoretical predictions would require more precise knowledge of $\lambda_D$, $Z_{wall}$ and $Z_{part}$ to be compared with the experimental values.

\subsection{Oscillations along x for engulfed particle in a tense membrane}\label{Appendoscill}

Upon tracking the motion of particles engulfed in a vesicle whose tension is controlled by a micropipette aspiration setup, it was found that the component of motion along the micropipette axis ($x$, as defined in Figure 6C) was affected by noise, leading to fluctuations in the MSD, as shown in Figure~\ref{oscill}A. This perturbation, when observed, systematically induced a peak in the position fluctuation spectrum along $x$ at a frequency of approximately $f \approx 60$ Hz, as shown in Figure~\ref{oscill}B. The exact origin of this perturbation could not be identified, but it can be attributed to the operation of another component on the optical table, leading to position fluctuations of the micropipette holder along $x$.Nevertheless, the presence of oscillations in the MSD curves along $x$ does not prevent the measurement of a relative decrease between more and less tense membrane states.\\

Additionally, the fact that the slope of the MSD curves along $x$ are consistently smaller than along $y$ points to a smaller diffusivity along $x$. This could stem from geometric effects arising from the fact that during the acquisition, the particle is not exactly at the bottom of the vesicle. This leads to the motion recorded along $x$ to also contain a small component along $z$ which can not be resolved in these 2-dimensional acquisitions. 

\subsection{Tension dependent diffusion of optically trapped wrapped particles}\label{optdiff}

In order to confirm the robustness of the tension-dependent diffusion of engulfed particles, we perform the experiment described in Figure \ref{tensiondeptrap}. By weakly trapping ($\kappa<10^{-6}$ N.m$^{-1}$) the engulfed particle to maintain it in focus together with the micropipette, we can perform the same experiment as the one performed in 6C. Again, during a single acquisition, we perform a sudden step-wise increase of the applied tension and subsequently track the motion of the particle in the optical trap. Figure \ref{tensiondeptrap} shows bright field images of the experiment and the plots of the two quantities used to extract the translational friction coefficient $\zeta_t$ and stiffness $\kappa$ for the motion parallel to the mother vesicle membrane (along the axis $y$, as defined in Figure 6C). Note that upon increasing the tension, the particle is slightly displaced from its equilibrium position in the trap along the direction perpendicular to the membrane. This results in a smaller effective trap stiffness along $y$ and explains the difference in plateau height between the red MSD curve (low tension) and black one (high tension). When fitting the data, $\kappa$ will therefore also be a fit parameter. Values extracted from fits of the MSD and PSD data are plot in Figure \ref{tensiondeptrap} \textit{(right)}. The measured values for the translational friction $\zeta_t$ confirm the existence of a tension-dependent dissipative process, as $\zeta_t$ shows a roughly twofold increase for a threefold tension increase.

\bibliography{suppbib}